# GENERALIZATION OF THE HYPERVIRIAL AND FEYNMAN-HELLMANN THEOREMS

**Teimuraz Nadareishvili[1,] and Anzor Khelashvili[2]**

[1] Iv. Javakhishvili Tbilisi State University Chavchavadze Ave. 3, 0162, Tbilisi, Georgia and Inst. of High Energy Physics, Iv. Javakhishvili Tbilisi State University, University Str. 9, 0109, Tbilisi, Georgia

[2] Inst. of High Energy Physics, Iv. Javakhishvili Tbilisi State University, University Str. 9, 0109, Tbilisi, Georgia and St.Andrea the First-called Georgian University of Patriarchy of Georgia, Chavchavadze Ave.53a, 0162, Tbilisi, Georgia.

*E-mail: teimuraz.nadareishvili@tsu.ge and anzor.khelashvili@tsu.ge*
Corresponding author. Phone: 011+995-98-54-47 ; E-mail :teimuraz.nadareishvili@tsu.ge

**Abstract.** Using well-known methods we generalize (hyper)virial theorems to case of singular potential. Discussion is carried on for most general second order differential equation, which involves all physically interesting cases, such as Schrödinger and two-body Klein-Gordon equations with singular potentials. Some physical consequences are discussed. The connection with Feynman-Hellmann like theorems are also considered and some relevant differences are underlined.

# 1. Introduction

Virial theorem has a wide application both in classical as well as in quantum mechanics. This theorem connects average values of kinetic and potential energies for the systems confined in limited areas. Moreover it allows making definite conclusions about some interesting problems without solving to equations of motion.

There are many generalizations of virial theorem, especially in relativistic quantum mechanics for investigation of bound states [1].

Recently much attention was devoted to singular potentials, namely, to potentials, behaving as $r^2 V(r) \to -V_0$ ; $(V_0 > 0)$ for $r \to 0$ in the Schrodinger equation, and as $rV = -V_0$ for $r \to 0$ in the Klein-Gordon and Dirac equations.

Such behaved potentials appear in large classes of physical problems. Particularly, in Calogero model [2], Coulomb or Hulthen potential in Klein-Gordon and Dirac equations [3], Black Hole theory [4] and etc. Virial-like theorems can make things clear while studying such problems.

Therefore it seems natural to make attempts for generalization of virial theorem for such (singular) potentials too.

The most general methods for obtaining various virial like theorems were developed in [5] by C. Quigg for regular potentials in the Schrodinger equation. The general character of these methods allows us to carry over singular potentials as well. It appears that formally the theorem almost keeps the form familiar for regular potentials with obvious differences in averaging procedure.

But the main differences are provided by additional solutions, which are the relevant property of singular potentials and is related to the necessity of self-adjoint extension (SAE).

This article is organized as follows:

First of all we remember the needed methods for deriving of virial-like theorems and apply them to general second order differential equation.

Consequences for regular potentials are reviewed and then the singular potentials are considered. It is shown, that there arise additional terms in the usual virial-like theorems, which depend on the additional solution for singular potentials. Some consequences of the new form of virial theorems are also considered.

After that the corresponding corrections to the Feynman-Hellmann theorem are discussed.

## II. Derivation of Hypervirial (generalized virial) Theorems

Let us consider the second order differential equation of most general form
$$R''(r) + G(r)R'(r) + L(r)R(r) = 0 \qquad (2.1)$$
Eexclusion of first derivative terms is always possible by using a suitable transformation [6]
$$u(r) = R(r) e^{\frac{1}{2}\int G dr} \qquad (2.2)$$
and for $u(r)$ function it follows the equation
$$u''(r) + D(r)u(r) = 0, \qquad (2.3)$$

where $D(r)$ is

$$D = H - \frac{1}{4}G^2 - \frac{1}{2}G' \tag{2.4}$$

But it turns out that the transformation (2.2) may be applied by some care. In particular we must distinguish two cases:

$$G(r) = \frac{2}{r} \tag{2.5}$$

and

$$G(r) \neq \frac{2}{r} \tag{2.6}$$

In case (2.5) equation (2.1) becomes

$$R''(r) + \frac{2}{r}R'(r) + L(r)R(r) = 0 \tag{2.7}$$

Central potential in three-dimensions will be important for us in the following. Exactly to equation (2.7) reduces the radial Schrodinger and the one- and two-body Klein-Gordon equations with $0 < r < \infty$. Even the one-dimensional case may be investigated on the same footing, as well, where $-\infty < x < \infty$. The transformation (2.2) gives

$$R(r) = \frac{u(r)}{r} \tag{2.8}$$

and the Eq. (2.7) takes a reduced form

$$u''(r) + L(r)u(r) = 0 \tag{2.9}$$

But a careful consideration [7-8] shows that after the substitution (2.8) into (2.7) there appears an additional contribution in the form of Dirac's delta-function, $-4\pi\delta^{(3)}(\mathbf{r})u(r)$, and for its removal one must impose the following behaviour of the reduced wave function at the origin

$$u(r) \underset{r \to 0}{\approx} r^N; \quad N = 1,2,3... \tag{2.10}$$

Moreover there appears that the satisfaction of this condition (2.10) is possible only for regular potentials. Therefore for singular potentials the substitition (2.2) does not remains the usual form of equation (2.9) and we are forced to explore directly the equation (2.7).

As regards of the second case (2.6), here the substitution (2.2) does not lead to such inconsistency. For example, if we take

$$G(r) = \frac{1}{r} \tag{2.11}$$

it gives

$$R(r) = \frac{u(r)}{\sqrt{r}} \tag{2.12}$$

Our aim is the derivation of the hypervirial theorem for the equation (2.7). Let rewrite it in the following manner

$$R''(r) + L(r)R(r) = -\frac{2}{r}R'(r) \tag{2.13}$$

This equation will be our starting point in this article.

Let us introduce an arbitrary three-times differentiable function $f(r)$, which will be restricted somehow in the following. The needed restrictions will follow from the requirements that arise below step by step. After multiplying the Eq.(2.13) on $f(r)$ and integrating obtained result in the interval $(0,\infty)$. We derive

$$-\int_0^\infty fR'R''r^2 dr = \int_0^\infty fLRR'r^2 dr + 2\int_0^\infty \frac{f}{r} R'^2 r^2 dr \qquad (2.14)$$

Let us mention that by using the following relations $R'R'' = \frac{1}{2}\left[(R')^2\right]'$ and $RR' = \frac{1}{2}\left[R^2\right]'$, one can perform the partial integration in (2.14)

$$-fr^2[R']^2\Big|_0^\infty + \int_0^\infty (fr^2)'[R']^2 dr = fr^2 LR^2\Big|_0^\infty - \int_0^\infty (fr^2)' LR^2 dr - \int_0^\infty fL'R^2 r^2 dr + 4\int_0^\infty frR'^2 dr$$
(2.15)

For bound states $R, R' \to 0$ at large distances and therefore one neglects contributions from upper bound in (2.15) if in addition $f$ and $L$ are to be restricted as follows

$$\lim_{r\to\infty} fr^2 R'^2 \to 0; \qquad \lim_{r\to\infty} fr^2 LR^2 \to 0 \qquad (2.16)$$

(For scattering problems $R, R'$ are not decaying functions and the conditions (2.16) may not take place, if we do not require it by special choice of $f$).

Therefore there remain integrated terms in (2.15) only at lower bound:

$$fr^2 R'^2\Big|_0 + \int_0^\infty (fr^2)' R'^2 dr = -fr^2 LR^2\Big|_0 - \int_0^\infty (fr^2)' LR^2 dr - \langle fL'\rangle + 4\int_0^\infty frR'^2 dr \qquad (2.17)$$

where $\langle\ \rangle$ denotes averaging by means of $R$ function

$$\langle fL\rangle = \int_0^\infty fLR^2 r^2 dr \qquad (2.18)$$

Now let us transform the second term in the RHS of (2.17)

$$\int_0^\infty (fr^2)' LR^2 dr = \int_0^\infty f'r^2 LR^2 dr + 2\int_0^\infty fLR^2 r\, dr = \langle fL\rangle + 2\left\langle \frac{f}{r}L\right\rangle \qquad (2/19)$$

and collect the last terms in both sides of (2.17)

$$I = \int_0^\infty (fr^2)' R'^2 dr - 4\int_0^\infty frR'^2 dr = \int_0^\infty Fr^2 R'^2 dr \qquad (2.20)$$

where

$$F = f' - \frac{2f}{r} \qquad (2.21)$$

Now perform a partial integration in the (2.20), using evident relation $(RR')' = R'R' + RR''$. It follows

$$I \equiv \int_0^\infty Fr^2 R'^2 dr = Fr^2 RR'\Big|_0^\infty - \int_0^\infty (Fr^2)' RR' dr - \int_0^\infty (Fr^2) RR'' dr \qquad (2.22)$$

For bound states the first term on RHS at the upper limit may be neglected, if

$$\lim_{r\to\infty} Fr^2 RR' \to 0 \qquad (2.23)$$

Now let us integrate the second term on RHS of (2.22)

$$I_1 = \int_0^\infty (Fr^2)' RR' dr = -\frac{1}{2}(F'r^2 + 2rF)R^2\Big|_0 - \frac{1}{2}\langle F''\rangle - 2\left\langle \frac{F'}{r}\right\rangle - \left\langle \frac{F}{r^2}\right\rangle \qquad (2.24)$$

Here we have taken into account that for bound states $F$ must be restricted as follows

$$\lim_{r\to\infty}(F'r^2 + 2rF)R^2 \to 0 \qquad (2.25)$$

In the same fashion the last integral in (2.22) takes the form

$$I_2 = \int_0^\infty (Fr^2) RR'' dr = FrR^2 \big|_0^\infty - \langle FL \rangle + \left\langle \frac{F'}{r} \right\rangle + \left\langle \frac{F}{r^2} \right\rangle \tag{2.26}$$

Here we made a further assumption that for bound states $F$ must be restricted as follows

$$\lim_{r \to \infty} FrR^2 \to 0 \tag{2.27}$$

Therefore, we have

$$I = \left[ -Fr^2 RR' + \frac{1}{2} F' r^2 R^2 \right]_0 + <FL> + \left\langle \frac{F'}{r} \right\rangle + \frac{1}{2} <F''> \tag{2.28}$$

At last, from (2.17), (2.21) and (2.28) for bound states we derive the following hypervirial theorem:

$$\left\{ f \left[ R^2 - r^2 RR'' + r^2 R'^2 \right] - f'rR[rR' + R] + \frac{1}{2} f'' r^2 R^2 \right\}_{r=0} = \tag{2.29}$$
$$= -2 <f'L> - <fL'> - \frac{1}{2} <f'''>$$

For scattering states (2.16), (2.23), (2.25) and (2.27) restrictions are not satisfied and instead of (2.29) we should have

$$-\left\{ f \left[ R^2 - r^2 RR'' + r^2 R'^2 \right] - f'rR[rR' + R] + \frac{1}{2} f'' r^2 R^2 \right\}_{r=\infty} + \tag{2.30}$$

$$+ \left\{ f \left[ R^2 - r^2 RR'' + r^2 R'^2 \right] - f'rR[rR' + R] + \frac{1}{2} f'' r^2 R^2 \right\}_{r=0} = -2 <f'L> - <fL'> - \frac{1}{2} <f'''>$$

Substituting here $R$ function at infinity, corresponding hypervirial theorem can be derived for scattering problems as well.

Now let us make some comments in connection to (2.29) about restrictions on $f$ :

(a) Because $< >$ means averaging by $R$ functions, $f$ must be such, that corresponding integrals do exist.

(b) When $f = r^q$ $(q \geq -2l)$, then (2.29) coincides with (2.27) from [10], in which only the Schrödinger equation was considered, i.e. when

$$L = 2m \left[ E - V - \frac{l(l+1)}{2mr^2} \right] \tag{2.31}$$

with regular $V$.

Let us note that the choice $f = r^q$ satisfies to (2.16), (2.23),(2.25) and (2.27) restrictions.

(c) The expression like (2.29) for arbitrary $f$ was derived in [11] for regular potentials only., and only for Schrodinger equation, as in [10].

### III. Some Applications of Hypervirial Theorem

Choosing $f$, one can obtain several interesting expressions from (2.29). Let us consider some of them.

Consider a particular case for $L(r)$ in (2.7)

$$L = A(r) - \frac{s(s+1)}{r^2}, \quad s \geq 0 \tag{3.1}$$

i.e. we separate a centrifugal term.

We use here a general notation $A(r)$ instead of (2.31) because a lot of physical equations reduce to the form, like (3.1), where potential participates in different manners.

It is necessary to make distinction between two cases: $\lim_{r\to 0} r^2 A(r) = 0$ (regular) and $\lim_{r\to 0} r^2 A(r) \neq 0$ (singular).

Consider each of them in detail:

(i) **regular case**, when

$$\lim_{r\to 0} r^2 A(r) = 0 \tag{3.2}$$

It is easy to guess, that only regular potentials

$$\lim_{r\to 0} r^2 V(r) = 0 \tag{3.3}$$

obey to (3.2) in case of Schrödinger equation (if we take $s = l$; $l = 0,1,2,...$), whereas, for example, for one- and two-particle Klein-Gordon equations the condition (3.2) will be satisfied if

$$\lim_{r\to 0} r V(r) = 0 \tag{3.4}$$

When (3.2) is satisfied it follows the following behavior of wave function at the origin

$$R \underset{r\to 0}{\approx} a_s r^s + b_s r^{-s-1} \tag{3.5}$$

The second term in (3.5) does not obey to the condition of hermiticity for Hamiltonian [12,13] and for the radial momentum operator $p_r = -i(\frac{\partial}{\partial r} + \frac{1}{r})$ [14], which is imposed on the wave function at origin

$$\lim_{r\to 0} r R(r) = 0 \tag{3.6}$$

Therefore it is neglected as a rule (see, any textbook in quantum mechanics). Then at small distances only the first term remains

$$R_s \approx a_s r^s \tag{3.7}$$

Substituting this into (2.29) one obtains

$$a_s^2 \left\{ r^{2s}\left[(s+1)f - (s+1)f'r + \frac{r^2}{2}f''\right]\right\}_{r=0} = -2<f'A> - <fA'> + \\ + 2s(s+1)<\frac{f'}{r^2} - \frac{f}{r^3}> - \frac{1}{2}<f'''> \tag{3.8}$$

Now consider a special form for $f$ [5]

$$f = r^q \tag{3.9}$$

We have

$$\left\{(s+1)(1-q) + \frac{1}{2}q(q-1)\right\} a_s^2 r^{q+2s}\Big|_{r=0} = -<2qr^{q-1}A + r^q A'> - \\ -\left[2s(s+1)(1-q) + \frac{1}{2}q(q-1)(q-2)\right]<r^{q-3}> \tag{3.10}$$

In order that the LHS of this expression be not diverging at $r = 0$, we must require

$$q \geq -2s \tag{3.11}$$

Therefore, (3.10) becomes

$$(2s+1)^2 a_s^2 \delta_{q,-2s} =$$
$$= - < 2qr^{q-1}A + r^q A' + \left[2s(s+1)(1-q) + \frac{1}{2}q(q-1)(q-2)\right]r^{q-3} > \tag{3.12}$$

It must be noted that (3.12) is a generalization of relation (2.30) from paper [5] for $L$ in form (3.1).

Let now consider various interesting values of $q$ in (3.12):

a) $q = 1$

Then it follows from (3.12) that
$$\langle 2A + rA' \rangle = 0 \tag{3.13}$$

In case of Schrodinger equation, when
$$A = 2m(E-V) \tag{3.14}$$

we derive
$$E = \left\langle V + \frac{1}{2}rV' \right\rangle \tag{3.15}$$

which is the usual virial theorem
$$<T> = \frac{1}{2}\langle rV' \rangle \tag{3.16}$$

b) $q = -2l$

Taking into account separability of total wave function
$$\psi(r,\theta,\phi) = R_{nl}(r)Y_{lm}(\theta,\phi) \tag{3.17}$$

we derive
$$(2l+1)^2 \left|R_{n,l}^{(l)}(0)\right|^2 = (l!)^2 < 4l\frac{A}{r^{2l+1}} - \frac{A'}{r^{2l}} >_{n,l} \tag{3.18}$$

Here $R_{n,l}^{(e)}(0)$ is the $l$-th order derivative of radial wave function at origin. Eq.(3.18) generalizes eq. (1.4) of [10] for Schrodinger equation, which has a form
$$(2l+1)^2 \left|R_{n,l}^{(e)}(0)\right|^2 = 2m(l!)^2 \left[\left\langle \frac{1}{r^l}\frac{dV}{dr}\right\rangle_l + 4l\left\langle \frac{E-V}{r^{2l+1}}\right\rangle\right] \tag{3.19}$$

c) q=0 or $f = const$

This case is well-known in the Schrodinger equation [5,11]. Now it follows from (2.29):
$$\left\{R^2 - r^2 RR'' + r^2 R'^2\right\}_{r=0} = - <L'> \tag{3.20}$$

or
$$(l+1)a_l^2 r^{2l}\Big|_{r=0} = - <A'(r)> - \left\langle \frac{2l(l+1)}{r^3}\right\rangle \tag{3.21}$$

If now we take $l = 0$, then
$$a_0^2 = [R_0(0)]^2 = - <A'(r)> \tag{3.22}$$

It generalizes eq. (39a) from [11] for arbitrary $A(r)$. When we take expression (3.14), then it follows from (3.22) the well-known relation
$$|\psi_0(0)|^2 = \frac{m}{2\pi}\left\langle \frac{dV}{dr}\right\rangle \tag{3.23}$$

In case $l \neq 0$, the LHS of (3.21) is zero and therefore we obtain
$$2l(l+1)\left\langle \frac{1}{r^3}\right\rangle = -\langle A' \rangle \tag{3.24}$$

which generalizes eq. (39b) from [11] for arbitrary $A(r)$. The relations (3.22) and (3.24) are formulated in terms of $A(r)$. Depending on equations of motion, the potential $V(r)$ appears in various forms and one must take care, which restrictions arise on potential $V(r)$.

d) $q \neq 0, 1, -2l$

Now we have

$$\left< 2qr^{q-1}A + r^q A' + \left[2l(l+1)(1-q) + \frac{1}{2}q(q-1)(q-2)\right]r^{q-3} \right> = 0 \qquad (3.25)$$

This expression allows us to connect average values of various degrees of $r$. For example, in Schrodinger equation we have

$$2Eq\langle r^{q-1}\rangle - 2q\langle r^{q-1}V\rangle - \langle r^q V'\rangle + \frac{(q-1)}{m}\left[\frac{q}{4}(q-2) - l(l+1)\right]\langle r^{q-3}\rangle = 0 \qquad (3.26)$$

For power-like potential, $V = V_0 r^n$ it follows from (3.26), that

$$2Eq\langle r^{q-1}\rangle - V_0(2q+n)\langle r^{q+n-1}\rangle + \frac{(q-1)}{m}\left[\frac{q}{4}(q-2) - l(l+1)\right]\langle r^{q-3}\rangle = 0 \qquad (3.27)$$

If $n = -1$, the well-known Kramer's formula [15] follows for the Coulomb potential $V = -\frac{\alpha}{r}$, (i.e. $V_0 = -\alpha; q = s+1$)

$$2E(s+1)\langle r^s\rangle + \alpha(2s+1)\langle r^{s-1}\rangle + \frac{s}{m}\left[\frac{s^2-1}{4} - l(l+1)\right]\langle r^{s-2}\rangle = 0 \qquad (3.28)$$

And when $n = 2$, the relation for isotropic harmonic oscillator $V = \frac{1}{2}\omega^2 r^2$ is derived [16]

$$2E(s+1)\langle r^s\rangle - \omega^2(s+2)\langle r^{s+2}\rangle + \frac{s}{m}\left[\frac{s^2-1}{4} - l(l+1)\right]\langle r^{s-2}\rangle = 0 \qquad (3.29)$$

Also it is possible to derive recurrence like relations between different powers of $r$ for various relativistic equations. Such relations have many applications in the diverse physical problems [17].

Let us note also that while we considered only regular potentials, we could work with the wave function $u(r) = rR(r)$ as well.

ii) **Singular case .** Now

$$\lim_{r \to 0} r^2 A(r) = -V_0; \qquad (V_0 > 0) \qquad (3.30)$$

As was shown in [18,19], in cases of Schrodinger and two equal mass particles' Klein-Gordon equations, besides the standard levels there appear additional levels as well, whose wave function behaves at small distances as

$$R_{st} \approx a_{st} r^{-\frac{1}{2}+P} \; ; \; R_{add} \approx a_{add} r^{-\frac{1}{2}-P} \qquad (3.31)$$

where, for example, in Schrodinger equation

$$P = \sqrt{(l+1/2)^2 - 2mV_0} > 0 \qquad (3.32)$$

while in the Klein-Gordon equation for two equal mass particles

$$P = \sqrt{(l+1/2)^2 - V_0^2/4} > 0 \qquad (3.33)$$

Likely it is possible to find P for a given L (see, above) for each relativistic equations. At the same time, as is indicated in [18,19], for the existence of additional levels following constraint must be satisfied

$$0 \leq P < 1/2 \tag{3.34}$$

which is expression of the fact that the condition (36) is satisfied.

Now if we take the wave function at small distances as general form [19]

$$R = a_{st} r^{-\frac{1}{2}+P} + a_{add} r^{-\frac{1}{2}-P} \tag{3.35}$$

and use (3.9) for $f$, then (2.29) gives

$$(1-q)(1/2 + P - q/2)a_{st}^2 \delta_{q,1-2p} + (1-q)(1/2 - P - q/2)a_{add}^2 \delta_{q,1+2P} +$$
$$+ \left[ (q-1)^2 - 4P^2 \right] a_{st} a_{add} \delta_{q,1} = \tag{3.36}$$
$$= -\left\langle 2qr^{q-1}A + r^q A' + \left[ 2l(l+1)(1-q) + \frac{q}{2}(q-1)(q-2) \right] \right\rangle \left\langle r^{q-3} \right\rangle$$

Here we must require that $q \geq 1 - 2P$.

Let consider various q-s in (3.36) as above.

a) $q = 1; \quad P \neq 0, 0 < P < 1/2$

Then from (3.36) follows

$$\left\langle 2A + rA' \right\rangle = 4P^2 a_{st} a_{add} \tag{3.37}$$

For the Schrodinger equation this means

$$E = \left\langle V + \frac{1}{2} rV' \right\rangle + \frac{P^2}{m} a_{st} a_{add} \tag{3.38}$$

Therefore, for singular potential the virial theorem differs from that of regular ones by the extra term

$$b = \frac{P^2}{m} a_{st} a_{add} \tag{3.39}$$

This term vanishes when we take only standard or only additional solutions.

*Comment*: Separate consideration needs the case $P = 0$. As is indicated in [19], we have in this case

$$R \underset{r \to 0}{\approx} a_{st} r^{-\frac{1}{2}} + a_{add} r^{-\frac{1}{2}} \ln r \tag{3.40}$$

Clearly $\lim_{r \to 0} rR(r) = 0$. Now instead of (3.36) it follows

$$\left\langle 2qr^{q-1}A + r^q A' \right\rangle - \left[ 2l(l+1)(1-q) + \frac{1}{2}q(q-1)(q-2) \right] \left\langle r^{q-3} \right\rangle = 0 \tag{3.41}$$

And virial theorem for Schrodinger theory takes the form

$$E = \left\langle V + \frac{1}{2} rV' \right\rangle \tag{3.42}$$

which is analogous to regular potential case, but difference appears in averaging by function with the behaviour (3.40).

For pure singular potential

$$V = -\frac{V_0}{r^2}; (V_0 > 0) \tag{3.43}$$

it follows from (3.37) that

$$E = \frac{P^2}{m} a_{st} a_{add} \qquad (3.44)$$

This is a single level, which appears in quantum mechanical consideration, when we retain the additional solution as necessary ingredient for providing a self-adjointness of Hamiltonian via self-adjoint extension (SAE) procedure [19].

This level disappears immediately as we neglect pure standard or pure additional solutions.

It is evident that the equality (3.37) is rather general relation and many physical consequences can be derived from it.

Consider, for example, two-particle Klein-Gordon equation with equal masses $m$:

$$R'' + \frac{2}{r} R' + \left[ \frac{V^2}{4} - \frac{MV}{2} + \frac{M^2}{4} - m^2 \right] R - \frac{l(l+1)}{r^2} R = 0; \qquad (3.45)$$

$M$ is a total mass of composite state.. Comparison to (2.7),(3.1) and (3.45) gives

$$A = \frac{V^2}{4} - \frac{MV}{2} + \frac{M^2}{4} - m^2 \qquad (3.46)$$

Using this in (3.37), we obtain

$$\left\langle \frac{V^2}{2} - MV + \frac{rV'}{2}(V - M) + \frac{M^2}{2} - 2m^2 \right\rangle = 4P^2 a_{st} a_{add} \qquad (3.47)$$

Let us now consider the following problem: Can two massive particles produce massless bound state in case of Coulomb potential (attraction or repulsion)? Existence of bound states for both cases is a consequence of the relativistic structure of Klein-Gordon equation, where for $M = 0$ there remains only $V^2$ in (3.45).This problem was considered in scientific literature [20].

For this aim one must take $M = 0$ in (3.47). We derive

$$\left\langle \frac{V^2}{2} + \frac{rV'}{2} V - 2m^2 \right\rangle = 4P^2 a_{st} a_{add} \qquad (3.48)$$

For Coulomb potential it follows

$$-m^2 = 2P^2 a_{st} a_{add} \qquad (3.49)$$

and we see that this problem has a positive answer only if both $a_{st} \neq 0$ and $a_{add} \neq 0$ (and $a_{st} a_{add} < 0$). Correctness of this result may be verified also by direct solution of the Klein-Gordon equation. Indeed, substituting $M = 0$ in (3.45), one finds

$$R'' + \frac{2}{r} R' + \left[ \frac{V^2}{4} - m^2 \right] R - \frac{l(l+1)}{r^2} R = 0; \qquad (3.50)$$

If we take here $V = \mp \frac{\alpha}{r}$ this equation becomes

$$R'' + \frac{2}{r} R' + \left[ -m^2 - \frac{P^2 - 1/4}{r^2} \right] R = 0 \qquad (3.51)$$

where $P$ is given by (3.33). Note that this equation coincides to the Schrodinger equation with the accuracy of notations. Therefore we can use the results of our paper [19] and write down the general solution derived there

$$R(r) = \sqrt{\frac{m}{r}} \{ A I_P(mr) + B I_{-P}(mr) \} \qquad (3.52)$$

where $I_P$ and $I_{-P}$ are the modified Bessel functions. We have the following behaviour at infinity

$$R(r) \underset{r \to \infty}{\approx} \frac{1}{\sqrt{2\pi r}} \{A + B\} e^{mr} \tag{3.53}$$

Requiring vanishing of $R(r)$ at infinity as for bound state solution we have to take

$$B = -A \tag{3.54}$$

Remembering the well-known relation

$$K_P(z) = \frac{\pi}{2 \sin P\pi} [I_{-P}(z) - I_P(z)] \tag{3.55}$$

our wave function takes the form

$$R = -A \frac{2}{\pi} \sqrt{\frac{m}{r}} \sin P\pi \cdot K_P(mr) \tag{3.56}$$

which is exponentially damping at infinity and in the interval $0 \leq P < 1/2$ satisfies to fundamental requirement (3.6). It is evident, that our solution is derived by the requirements

$$A \neq 0; \quad B \neq 0 \tag{3.57}$$

which means, that $M = 0$ state can be derived only by SAE procedure. We see that explicit solution of Klein-Gordon equation repeats the conclusion, derived by Virial theorem.

One important remark is in order: W. Krolikowski [20] derived the same solution for $l = 0$ state only. It is true, because $K_P(z)$ is the only Bessel function, which behaves in a needed fashion at infinity (vanishes!).It appears that a massless bound state for Coulomb potential may be constructed from 2 massive particle in nonzero orbital momentum states also, $l \neq 0$ [18].But SAE procedure is necessary.

Owing to the fact, that repulsive case also forms a massless bound state, we conclude, that the following alternatives take place:

(i) Those values of SAE parameter $\tau = \frac{a_{add}}{a_{st}}$, when this strange fact occurs, must be

 deflected in order to suppress such unphysical results.
(ii) We must recognize, that the SAE procedure produces an *effective attraction*, which may be seen from equation (3.50), where the factor $(P^2 - 1/4)$ has a negative sign in area (3.34) and gives a quantum anticentrifugal potential, which is attractive [19].
(iii) It is not excepted that such unphysical fact is a pathology of the Klein-Gordon equation. For example if we reverse the problem and ask if two massless particles can compose a massive bound state in Coulomb field, we can easily see that (3.47) would give a positive answer in case of Coulomb repulsion, but not for attraction .

 b) Cases $q = 1 \mp 2P$ and $q \neq 0, 1, -2l$ may be discussed in full analogy. One derives some recurrence like relations between average values of various powers of $r$.

### IV. Generalization of the Feynman - Hellmann Theorem

It is known that the Feynman-Hellmann (FH) [21 -22] and generalized FH [23-24] theorems are closely related to the hypervirial theorems.
 FH like theorems connect average values of energy derivatives by some parameters to those of Hamiltonians.
We want to take attention to the fact, that for singular potentials in Schrödinger equation, when SAE is necessary, FH theorem also should be modified.
 Indeed, in a traditional way [23], we consider a wave equation of the form

$$F(E,\lambda)\psi = 0 \tag{4.1}$$

where $\psi$ is the wave function of a bound state with energy E, which depends on the parameter $\lambda$. Then we can write

$$\frac{\partial}{\partial\lambda}\langle\psi|F|\psi\rangle = \left\langle\frac{\partial\psi}{\partial\lambda}\Big|F\Big|\psi\right\rangle + \left\langle\psi\Big|\frac{\partial F}{\partial E}\frac{\partial E}{\partial\lambda} + \frac{\partial F}{\partial\lambda}\Big|\psi\right\rangle + \left\langle\psi\Big|F\Big|\frac{\partial\psi}{\partial\lambda}\right\rangle = 0 \tag{4.2}$$

If F has the property that (this property is fulfilled in the regular (3.2) case!)

$$\left\langle\psi\Big|F\Big|\frac{\partial\psi}{\partial\lambda}\right\rangle = \left\langle F\psi\Big|\frac{\partial\psi}{\partial\lambda}\right\rangle \tag{4.3}$$

then in view of (4.1), from (4.2) we obtain

$$\frac{\partial E}{\partial\lambda} = -\frac{\left\langle\psi\Big|\frac{\partial F}{\partial\lambda}\Big|\psi\right\rangle}{\left\langle\psi\Big|\frac{\partial F}{\partial E}\Big|\psi\right\rangle} \tag{4.4}$$

If $F = H - E$, then (4.4) reduces to the usual Feynman – Hellmann theorem [23-24]

$$\frac{\partial E}{\partial\lambda} = \left\langle\psi\Big|\frac{\partial\hat{H}}{\partial\lambda}\Big|\psi\right\rangle \tag{4.5}$$

Now from (2.7), (3.1) and (4.1) we have

$$F = \frac{d^2}{dr^2} + \frac{2}{r}\frac{d}{dr} + A(r) - \frac{l(l+1)}{r^2} \tag{4.6}$$

and in the singular (3.30) case, instead of (4.3) we obtain additional term on the right side

$$\left\langle R_n\Big|F\Big|\frac{\partial R_n}{\partial\lambda}\right\rangle = \left\langle FR_n\Big|\frac{\partial R_n}{\partial\lambda}\right\rangle + \lim_{r\to 0}\left[rR_n(r,\lambda)\frac{d}{dr}\frac{\partial rR_n(r,\lambda)}{\partial\lambda} - \frac{\partial rR_n(r,\lambda)}{\partial\lambda}\frac{drR_n(r,\lambda)}{dr}\right] \tag{4.7}$$

and it follows

$$\frac{\partial E}{\partial\lambda} = -\frac{\left\langle\psi\Big|\frac{\partial F}{\partial\lambda}\Big|\psi\right\rangle + B}{\left\langle\psi\Big|\frac{\partial F}{\partial E}\Big|\psi\right\rangle} \tag{4.8}$$

where

$$B = \lim_{r\to 0}\left[rR_n(r,\lambda)\frac{d}{dr}\frac{\partial rR_n(r,\lambda)}{\partial\lambda} - \frac{\partial rR_n(r,\lambda)}{\partial\lambda}\frac{drR_n(r,\lambda)}{dr}\right] \tag{4.9}$$

We see that FH theorem is modified as well.

This happens because $F$ is not a self-adjoint operator in singular case (3.30) and the fundamental relation

$$\langle\psi|F|\varphi\rangle = \langle F\psi|\varphi\rangle \tag{4.10}$$

does not takes place. Therefore the SAE procedure is necessary.
Inserting (3.35) we obtain

$$B = -2P\left[a_{n,st}\frac{\partial a_{n,add}}{\partial\lambda} - a_{n,add}\frac{\partial a_{n,st}}{\partial\lambda}\right] + \frac{dP}{d\lambda}\lim_{r\to 0}\{4Pa_{n,st}a_{n,add}\ln r - a_{n,add}^2 r^{-2P}\} \tag{4.11}$$

Consider some consequences for Schrodinger and one body Klein-Gordon equation.

a) For Schrodinger equation P is given by (3.32) and for singular

$$\lim_{r \to 0} r^2 V = -V_0 \quad (V_0 > 0) \tag{4.12}$$

potential from (4.7) we obtain

$$\frac{\partial E_n}{\partial \lambda} = \left\langle R_n \left| \frac{\partial \hat{H}_r}{\partial \lambda} \right| R_n \right\rangle + \frac{P}{m}\left[ a_{n,st} \frac{\partial a_{n,add}}{\partial \lambda} - a_{n,add} \frac{\partial a_{n,st}}{\partial \lambda} \right] - \\ - \frac{1}{2m} \frac{dP}{d\lambda} \lim_{r \to 0} \left\{ 4 P a_{n,st} a_{n,add} r^{-1} \ln r - a_{n,fdd}^2 r^{-1-2P} \right\} \tag{4.13}$$

We see that the last parenthesis of eq. (4.13) is divergent expression at the origin, except the regular case or when $a_{add} = 0$. Therefore only for $\frac{\partial P}{\partial \lambda} = 0$ has this expression a viable sense, i.e. when we choose $\lambda \neq m, V_0$ or $l$.

So when SAE procedure (which is necessary in a singular potential case) is not used, the FH theorem takes usual form (4.5).

When P does not depend on $\lambda$ ($\lambda \neq m, V_0$ or $l$), there remains only the first row in (4.13)

$$\frac{\partial E_n}{\partial \lambda} = \left\langle R_n \left| \frac{\partial \hat{H}_r}{\partial \lambda} \right| R_n \right\rangle + \frac{P}{m}\left[ a_{n,st} \frac{\partial a_{n,add}}{\partial \lambda} - a_{n,add} \frac{\partial a_{n,st}}{\partial \lambda} \right] \tag{4.14}$$

In particular case, when $P = 0$, calculations must be performed by function (3.40). In this case singularities from (4.13) disappear. We find

$$\frac{\partial E_n}{\partial \lambda} = \left\langle R_n \left| \frac{\partial \hat{H}_r}{\partial \lambda} \right| R_n \right\rangle + \frac{1}{2m}\left[ a_{n,st} \frac{\partial a_{n,add}}{\partial \lambda} - a_{n,add} \frac{\partial a_{n,st}}{\partial \lambda} \right] \tag{4.15}$$

a) For one body Klein - Gordon equation

$$R'' + \frac{2}{r} R' + \left[ (E-V)^2 - m^2 - \frac{l(l+1)}{r^2} \right] R = 0 \tag{4.16}$$

We obtain for $\lambda \neq V_0$ or $l$

$$\frac{\partial E}{\partial \lambda} = \frac{1}{E - \langle V \rangle} \left\{ \left\langle (E-V) \frac{\partial V}{\partial \lambda} \right\rangle - \frac{B}{2} \right\} \tag{4.17}$$

And for $\lambda = m$ we get

$$\frac{\partial E}{\partial m} = \frac{m}{E - \langle V \rangle} + P\left\{ a_{st} \frac{\partial a_{add}}{\partial m} - a_{add} \frac{\partial a_{st}}{\partial m} \right\} \tag{4.18}$$

where

$$P = \sqrt{(l+1/2)^2 - V_0^2} > 0 \tag{4.19}$$

The main result here is that in case of singular potentials Feynman-Hellmann theorem has to be modified.

## V. Conclusions

In this article we consider problems, related to the singular potentials in light of hypervirial and FH theorems. Main results can be summarized as follows:
1. We have derived a hypervirial theorem for general second order differential equation.
2. For regular potentials we have generalized known results concerning the Schrodinger equation ( virial theorem, wave function and its derivatives at origin, recurrence relations between average values of different powers of $r$ )
3. We obtain virial theorem for singular potential, by means of which some physical results are derived (existence of one level for pure $r^{-2}$ potential, possibility of having massless bound state for repulsive and attractive Coulomb potential in the two-body Klein – Gordon equation).
4. We have derived the modification of Feynman-Hellmann theorem for singular potential, when the SAE procedure is necessary.

**Acknowledgments**

The authors thank A.Kvinikhidze M.Nioradze and participants of seminars at Iv. Javakhishvili Tbilisi State University, for many valuable comments and discussions. The designated project has been fulfilled by financial support of the Georgian National Science Foundation (Grants DI/13/02 and FR/11/24)).